\documentstyle[twoside,fleqn,espcrc2]{article}

\newcommand{\AmS}{{\protect\the\textfont2
  A\kern-.1667em\lower.5ex\hbox{M}\kern-.125emS}}

\def\NPB#1#2#3{Nucl. Phys. B {\bf #1}, #3 (19#2)}
\def\PLB#1#2#3{Phys. Lett. B {\bf #1}, #3 (19#2)}

\def\PRD#1#2#3{Phys. Rev. D {\bf #1}, #3 (19#2)}
\def\PRL#1#2#3{Phys. Rev. Lett. {\bf#1}, #3 (19#2)}

\begin{document}


\title{Dark-matter in gravity-mediated 
     supersymmetry breaking}

\author{James D.\ Wells\address{Stanford Linear Accelerator Center,
 Stanford, California 94309 }
        \thanks{Work supported by the Department of
 Energy, contract DE-AC03-76SF00515. SLAC-PUB-7605.}}
       

\begin{abstract}
In $R$ parity conserving supersymmetric theories, the lightest
superpartner (LSP) is stable.  The LSPs may comprise a large
fraction of the energy density of the current universe, which would
lead to dramatic astrophysical consequences.  In this talk, I will
discuss some of the main points we have learned about supersymmetric
models
from relic abundance considerations of the LSP.
\end{abstract}

\maketitle

\section{Introduction}

Astrophysicists have long been telling us that the universe is 
mostly made up of dark matter.  Modern analyses, which take into
account
the subtleties of large scale structure formation, big bang 
nucleosynthesis, and
observations
of how galaxies rotate, have largely condensed
to one common conclusion: 
most of
the dark matter is probably stable weakly interacting massive
particles (WIMPs)~\cite{r1}.  

While the astrophysics community was coming to
grips with the properties of the dark matter, the supersymmetry
community was working on its own problems.  In the early 1980's
it was first recognized that the proton would decay too quickly
if all allowed gauge invariant renormalizable operators in
the superpotential had order one strength.  A discrete symmetry
was postulated which eradicated these unwanted baryon and lepton
violating interactions~\cite{r2}.  The postulated symmetry, $R$-parity,
gave a positive charge to all standard model states, and negative
charge to all superpartner states.  The name ``$R$-parity''
is a somewhat unfortunate name since the symmetry is not intrinsically
$R$-symmetric but rather an ordinary global discrete symmetry valid for
the superfields (all matter fields are negatively charged, and all
Higgs fields are positively charged).  (A better name perhaps would
have been ``Matter parity'', however {\it stare decisis} dictates
that we continue using $R$-parity.)

It was soon realized that $R$-parity conservation also implies that
the lightest superpartner is stable.  A short cognitive leap from
this understanding is the realization that there might be many of
these stable particles still hanging around the universe.  In 1977
several authors demonstrated how to estimate the relic abundance of
stable particles (stable neutral leptons were of primary interest then)
which were in thermal equilibrium with the photon
bath in the early universe~\cite{r3}.  The connection between that work
and the existence of a stable neutral supersymmetric particle
was quick.  Weinberg~\cite{r4} was one of the first (in print) to
make the connection when he made the following
ancillary comment in his gaugino masses paper: ``... there is no clear
conflict [of the photino's mass] with cosmology, and we have a hint
that photinos may provide an important 'dark' contribution to
the cosmic mass density.''  Soon after that Goldberg~\cite{r5} presented
his paper on neutralino relic abundance.

\section{$R$-parity}

Many useful papers have followed Goldberg's work, and many important
points relating to neutralino dark matter have been discovered,
refined and debated.  One important debate is the origin of $R$-parity.
$R$-parity is overkill since
it banishes both baryon and lepton number violating operators, when
in reality only one need be erased.  (Proton decay, of course, proceeds via
baryon and lepton number violation.)  Therefore $R$ parity is not
unique in stabilizing the proton.  Other discrete symmetries such
as ``baryon parity'' can do the job as effectively~\cite{r6}.  Furthermore,
the applicability of global symmetries has always been debated.
Detractors have several arguments ranging from 
``why should global symmetries exist'', 
to the catalepsy inducing ``worm holes violate global symmetries.''
Of course no one would argue with accidental global symmetries which
are based on particle content and {\it gauge} symmetries.   Several
authors have focused on the gauge symmetry part and have discovered
that continuous gauge symmetries can spontaneously break down to 
a discrete gauge symmetry~\cite{r7}.  

From the low energy perspective the only
difference between a discrete gauge symmetry and a 
discrete global symmetry is
the former must identically solve a set of discrete anomaly
diophantine equations.  It turns out that $R$-parity is the only
$Z_2$ discrete symmetry which is anomaly free given the minimal supersymmetric
particle spectrum~\cite{r8}.  Practitioners devoted to simplicity and
the pre\"eminence of gauge symmetries cannot help but be impressed
with $R$-parity as the solution to the proton stability question.
The work-horse continuous gauge symmetry which could give
rise to $R$-parity
is $U(1)_{B-L}$.  Any group which contains $U(1)_{B-L}$ 
has the
potential to spontaneously break down to the standard model plus
$R$-parity as long as the order parameter is of the right conjugacy
class~\cite{r9}.  Candidate groups include the well-motivated
$SO(10)$, $SU(4)$, $SU(2)_L\times SU(2)_R$, and more.  Nature could
well give us $R$-parity conservation from these higher rank groups.
More progress will surely come to light on how motivated 
$R$-parity is for the low energy theory.  Without ever considering
the positive ramifications of supersymmetric dark matter, $R$-parity
still survives as a likely candidate symmetry to protect the proton
from decaying too quickly.

\section{What is the LSP?}

I'll assume $R$-parity conservation for the rest of the talk, and
therefore the LSP is stable.  What's the LSP?  This review is on
gravity mediated theories, however it should be pointed out that
if supersymmetry is broken at low scales then the gravitino
could be the LSP~\cite{r10}.  Depending on its mass it too could be a dark
matter candidate but it is warm dark matter rather than the
more preferred cold dark matter which cosmologists find
so appealing.  Nevertheless, in low-energy breaking supersymmetry
theories cold dark matter candidates can be found such as the
messengers in theories which communicate the supersymmetry breaking
via gauge interactions~\cite{r11}.  I won't discuss such theories further, and
will only focus on the gravity mediated case.  Part of what I am implying by
``gravity mediated'' is the assumption
that the gravitino is heavy and irrelevant for our discussion
and that no other states exist near the weak scale except MSSM states.

If we just write down the most general softly broken supersymmetric
lagrangian with standard model
gauge symmetries and $R$-parity conservation, we find that there are
over one hundred free parameters corresponding to the masses, flavor 
mixing angles, and CP violating angles.  Numerous simplifications
are often imposed such as universality among scalar masses and among
gaugino masses at the high scale, flavor angle alignment with the standard
model CKM angles, and zero CP violating phases beyond the single phase
in the CKM matrix.  Not all of these restrictive assumptions
are necessary simultaneously in many of the points that I will outline
below.  Unless otherwise stated, I will always assume that gaugino mass
unification occurs at the high scale.  In most cases 
it is straightforward to
generalize results when the simplifying assumptions are abandoned.

The dark matter is probably not charged~\cite{r12}, so that leaves us with two
possibilities for the dark matter:  a sneutrino or the lightest
neutralino.  There are several problems with the sneutrino as
a dark matter candidate.  First, it interacts rather efficiently with
ordinary matter, and if it constitutes much of the dark matter 
in our galactic halo then it should have already been detected
up to the TeV mass range~\cite{r13}.  This covers a lot of ground
in the sneutrino parameter space. More importantly, such high supersymmetry
masses call into question 
the natural solution to electroweak symmetry breaking provided by
supersymmetric theories. Second, renormalization group analyses 
demonstrate that there always exists at least one neutralino lighter
than the sneutrino if the sneutrino mass is above 80~GeV~\cite{r14}.  This
statement is valid for any positive intrinsic soft supersymmetry breaking
scalar masses at the high scale.  At such low mass, the sneutrinos could
not provide an interesting amount of dark matter (they annihilate very
efficiently through the $Z$ boson).  Being $SU(2)$ partners with
left-handed charged leptons, signatures at FNAL and LEPII should
rule out the entire region below 80 GeV from slepton production and
decay.  Therefore, it is likely that sneutrinos are not the cold
dark matter of the universe.

On the other hand, neutralinos provide a very nice dark matter candidate.
For one, they usually come out the lightest particle given a survey
over minimal model boundary conditions at the high scale~\cite{r15}.  Second, the
composition is almost pure bino, which means that it doesn't couple
at full $SU(2)$ strength to the $Z$ boson.  The bino is almost pure bino
for several reasons.  The renormalization group equations for gauginos
dictate that the lightest gaugino at the weak scale be the bino.
It is a factor of two lighter than the wino.  The neutralino is a mixture
of the bino, wino, and two higgsino states which scale roughly with
the $\mu$ parameter.  The $\mu$ parameter is a mass parameter in the 
Higgs potential that must be at precisely the correct value such
that at the minimum of the potential $m_Z=91.19$~GeV.  The minimization
conditions depend on the values of $\tan\beta$, $m_{H_u}^2$ and
$m_{H_d}^2$.  Usually, $m^2_{H_u}$ gets renormalized to rather large
negative values scaling like the heavy top squark mass.  To compensate
for this large negative value, the $\mu^2$ term in the potential must
be large and positive, and it is typical that $|\mu |$ is 
substantially larger than the bino mass parameter.  Therefore,
a state which is mostly bino is the lightest neutralino.  Of course, this
is a conclusion based on the minimal model, but it has wide range
of applicability in non-minimal models as well.  I will briefly discuss
later the implications of non-minimality.

\section{Mass limits from relic abundance}

A mostly bino LSP is highly desirable~\cite{r16}, 
since, as noted above, it doesn't
interact well with the $Z$ boson.  Therefore, annihilations
of two binos into the $Z$ boson are not efficient and the binos
fall out of equilibrium faster, having a rather large relic number
density.  It is of course general for any WIMP; if it annihilates efficiently
then there are few leftover today.  A non-relativistic particle's
number density falls rapidly if it continues to stay in equilibrium
with the photons.  However, once it freezes out of equilibrium (interactions
can't keep up with the expansion of the universe) then it no longer
tracks the equilibrium number density all the way to zero.
In fact, the relic density scales inversely proportional to its annihilation
rate.  Since by dimensional analysis the annihilation rate
must scales as $1/m^2_{susy}$, and therefore the relic abundance
scales as $m^2_{susy}$.  It should be no surprise then that as the
supersymmetry breaking masses go higher and higher then the relic
abundance gets too large.  (That is the mass density calculation is
incompatible with the Hubble constant and current age of the universe.)
Therefore, there must be an upper limit to $m_{susy}$.

This upper limit can be illustrated nicely in the case of a pure bino.
For this case we assume, somewhat realistically, that the only other
relevant light particles in the spectrum are the right-handed sleptons.
In this case, $m_{susy}$ of the previous paragraph becomes a complicated
function of the slepton mass and the bino mass.  Drees and Nojiri~\cite{r17} 
showed that in this model the lightest neutralino and right handed sleptons
had to be below 200 GeV in order to not become incongruous with cosmological
data.  This remarkable result places an upper limit on two superpartner
masses from physical principles alone.  In other words, no 
insubstantial finetuning criteria need be placed on 
the electroweak symmetry breaking equations to obtain upper limits
on the superpartner masses.

It is probably not realistic to assume that nature agrees with a pure
bino LSP model.  More detailed model analyses which solve the electroweak
symmetry breaking equations and all the renormalization group equations
of the minimal model (perhaps also not realistic) maintain the general
result that superpartner masses are cutoff by relic abundance requirements.
In Fig.~6 of ref.~\cite{r15} one can see the effect of the relic abundance
constraint on the superpartner spectrum.  The effect is most easily
visualized by fixing the gaugino masses to a particular value and then
increasing the scalar masses to higher and higher values.  Since the
(mostly) bino of the minimal model does not couple well with the $Z$,
its main interactions are by $t$-channel slepton and squark exchange.
As these scalar masses get higher the annihilation rate decreases and
the relic abundance increases.  At sufficiently high scalar mass the relic
abundance becomes unacceptably large, indicating a cutoff in how
the scalar masses can go.  On the other hand, if the scalar masses
are fixed in value, and the gaugino masses are raised, other catastrophic
problems arise.  For example, the LSP might become charged (usually
the right-handed slepton), or the electroweak symmetry breaking
equations have no correct solution.  In any event, there is a cutoff
in the superpartner masses.  
The fact that cosmological arguments such as the above can yield upper
limits to the superpartner spectrum is one of the most important things
we have learned.

There are many ways to study how non-minimality affects dark
matter viability, or how dark matter viability affects non-minimal
models.  Certainly it is important to study how non-universal 
scalar masses interplay with dark matter.  This is done by
Arnowitt in these proceedings~\cite{r19}.  Other ideas include playing around
with the neutralino mass matrix to see if other equally 
attractive dark matter particles come out.  
A theme permeating all these types of analyses is the requirement
that the lightest neutralino not interact with the $Z$.  Both the
photino and the bino, long-studied dark matter candidates, satisfy
this requirement.  Other possibilities include the zino~\cite{r20}, 
sterile neutralino~\cite{r21} and the symmetric higgsino~\cite{r22,r23}.  
By sterile neutralino
I mean an additional degree of freedom in the neutralino mass matrix
(such as the superpartner to a new $Z'$ gauge boson or singlet Higgs).  

\section{Higgsino dark matter}

Realization that a weak-scale higgsino could be a legitimate dark matter
particle is a rather recent development.  
One way to obtain an higgsino as the lightest neutralino is
to make $|\mu |$ much less than the gaugino parameters in
the neutralino mass matrix.  A very low value of $\mu$ will 
create a roughly degenerate triplet of higgsinos.  The charged
higgsino and the neutral higgsinos can all coannihilate together
with full $SU(2)$ strength, allowing the LSP to stay in thermal
contact with the photons more effectively, thereby lowering the
relic abundance of the higgsino LSP to an insignificant level.
These coannihilation channels are often cited as the 
reason why higgsinos are not viable dark matter candidates.
This claim is true in general, but there are two specific cases
that I would like to summarize below that allow the higgsino
to be a good dark matter candidate.

Drees {\it et al.\ } have pointed out that potentially large 
one-loop splittings among
the higgsinos can render the coannihilations less relevant~\cite{r23}.
Under some conditions with light top squark masses,
one-loop corrections to the neutralino mass matrix will split
the otherwise degenerate higgsinos.  If the mass difference can
be more than about 5\% of the LSP mass, then the LSP will
decouple from the photons alone and not with its other higgsino
partners, thereby increasing its relic abundance.  

Another possibility~\cite{r22} relating to a higgsino LSP is to 
set equal the bino and wino mass to approximately $m_Z$.  Then
set the $-\mu$ term to less than $m_W$. This non-universality
among the gauginos and particular choice for the higgsino
mass parameter, produces a light higgsino with mass
approximately equal to $\mu$, a photino with mass at about
$m_Z$, and the rest of the neutralinos and both charginos with mass
above $m_W$.  There are no coannihilation channels to worry about
with this higgsino dark matter candidate since it no other
chargino or neutralino mass is near it.  
The value of $\tan\beta$ is also required to be
near one so that the lightest neutralino is an almost pure
symmetric combination of $\tilde H_u$ and $\tilde H_d$ higgsino
states.  The exactly symmetric combination does not couple to
$Z$ boson.  The annihilation cross section near $\tan\beta \sim 1$
is proportional to $\cos^2 2\beta$.  The relic abundance scales
inversely proportional to this, and so the nearly
symmetric higgsino in this case is a very good dark matter candidate.
Note that there are no $t$-channel slepton or squark diagrams since
higgsinos couple to squark proportional to the fermion mass.  Because
the higgsino mass is below $m_W$, the
top quark final state is kinematically inaccessible, and so the
large top Yukawa cannot play a direct role in the higgsino annihilations.

This non-minimal higgsino dark matter candidate described in the previous
paragraph was motivated by the $e^+e^-\gamma\gamma$ event 
reported by the CDF collaboration at Fermilab~\cite{r26}.  The non-minimal 
parameters
which leads to a radiative decay of the second lightest neutralino
(photino) into the lightest neutralino (symmetric higgsino) and photon
also miraculously yield a model with a good higgsino dark matter
candidate.

It should be noted that LEPII should be able to find the higgsino
dark matter candidate.  This is true because the higgsino mass must
be below $m_W$ in order to be a viable dark matter particle, and other
charginos and neutralinos should have masses which hover around
$m_W$.  If
its mass is higher than $m_W$ then the $W^+W^-$ annihilation channel
opens up at full $SU(2)$ strength with no suppressions, leaving the
density of relic higgsinos too small to be significant.  (It is also possible
that the higgsino could be in the multiple TeV region where it would
start to again perhaps become a good dark matter candidate.)
This is the reason that LEPII will be able to verify or rule out
the light higgsino dark matter idea after it has taken data
above 190~GeV.

\section{Conclusion}

Finally, there is still much to be done
both experimentally and theoretically on the dark matter question.
Experimentally, table top experiments, neutrino telescopes, 
cosmic ray detectors, etc., could all start becoming sensitive to
supersymmetric neutralino relics in the next few years~\cite{r27}.  Currently,
they typically fall a few orders of magnitude away from the 
expected signal.  There is also more work to be done in the theoretical
community.  For example, demonstrating how $R$-parity can arise naturally
from a string theory or from an elegant grand unified theory would
be an interesting development which should predict
ramifications for
other phenomenological aspects of the model (extra $Z'$, or 
exotic D-term effects).  High energy colliders in the near future
might be the first to detect the dark matter from decays of 
superpartners.  However, to demonstrate that stable particles on
detector time scales are real dark matter candidates that
live at least as long as the age of the universe requires
experiments specifically devoted to the task.

\end{document}